\documentclass[journal]{IEEEtran}

\usepackage{float}
\usepackage{epsfig}

\usepackage{graphicx}
\usepackage{caption}
\usepackage{amsmath}
\usepackage{algorithm}
\usepackage{algorithmicx}
\usepackage{algpseudocode}

\usepackage[caption=false]{subfig}

\newcommand{\RNum}[1]{\uppercase\expandafter{\romannumeral #1\relax}}

\begin{document}
%
\title{Estimation of Parameters in Avian Movement Models}
%
%
%

\author{Hua Bai}

\markboth{Electrical and Computer Engineering Department}%
{Shell \MakeLowercase{\textit{et al.}}: Bare Demo of IEEEtran.cls for IEEE Journals}
%

\maketitle

\begin{abstract}
The knowledge of the movement of animals is important and necessary for ecologists to do further analysis such as exploring the animal migration route. 
A novel method which is based on the state space modeling has been proposed to track the bird, where the VHF transmitter is attached to the bird to emit the signal and several towers with antenna arrays installed on its top are built to receive the signal.    
The method consists of two parts, the first one is called movement model which accounts for prediction of the dynamic movement of the target, and the second part is the measurement model which links the target's state variables to the available measurements data, the measurement includes the time when the signal was detected, the ID of the antenna array which detected the signal and integers between 0 and 255, the integers are proportional to the strength of received signal.
The extended Kalman filter is then applied to estimate the location of the target with combing the movement model and measurement model.
In the movement model, several parameters with positive values are deployed to define the change of the state variables with time, these parameters reflect the relationship of the state variables at current time and next time.
In this paper, a method based on the maximum likelihood estimation is proposed to estimate the appropriate values for these unknown constant variables with given measurement data, and a kite is applied to demonstrate the validity of the proposed method.
Furthermore, the unscented transformation is applied in Kalman filter to achieve more accurate estimation of the target's states.    
\end{abstract}


\IEEEpeerreviewmaketitle

\section{Introduction}
\IEEEPARstart{T}{he} knowledge of animals movements are important to ecologists, in order to analyze the certain species behavior and evolution in this area, such as the change of their habitat area, the use of the habitat, their behavior in migration and so on. The state space method \cite{Durbin1} \cite{Patterson1} has been demonstrated to be a good option in tracking the target in \cite{janaswamy_bird_1}.

The approach in \cite{janaswamy_bird_1} contains two parts, the state model which includes the spatial position and the velocity of the target. In the horizontal ($xy$) plane, the motion is governed by
\begin{align} \label{xstate}
 \dot{p_{x}}(t)&=v_{x}(t),\\
 \dot{v_{x}}(t)&=-\beta_{x}v_{x}(t)+\sigma_{x} N(t),
\end{align}
where $\dot{p_{x}}(t)$ and $\dot{v_{x}}(t)$ are the first derivative with respect to $t$, $\beta_{x}$ and $\sigma_{x}$ are two scalar constants, $N(t)$ is the white noise process which is independent of state variables.
In the vertical ($z$) plane, the governing equation is 
\begin{equation}
 \dot{p_{z}}(t)=-\beta_{z}p_{z}(t)+\sigma_{z}N(t),\\
\end{equation}
the transition equations for the entire system can be written in the matrix form
\begin{equation} \label{state1}
 \bar{X}(t+\Delta t)=\begin{bmatrix}H_{x} & 0 & 0 \\ 0& H_{y} & 0\\ 0& 0 &H_{z} \end{bmatrix} \bar{X}(t)+\begin{bmatrix}\nu_{x} & 0 & 0 \\ 0& \nu_{y} & 0\\ 0& 0 &\nu_{z} \end{bmatrix},
 \end{equation}
where $\bar{X}(t)=[p_{x}(t),v_{x}(t),p_{y}(t),v_{y}(t),p_{z}(t)]'$,the system matrix is 
\begin{equation}\label{state2}
H_{x}=H_{y}=\begin{bmatrix}1 &  \frac{1-e^{-\beta_{x}(\beta_{y})\Delta t}}{\beta_{x}(\beta_{y})}  \\ 0& e^{-\beta_{x}(\beta_{y})\Delta t} \end{bmatrix}, H_{z}=e^{-\beta_{z}\Delta t},
\end{equation}
$\nu_{x}$, $\nu_{y}$ and $\nu_{z}$ are subject to the normal distribution with zero mean, the covariance is  
\begin{equation} \label{state3}
   Q_{x}= \begin{bmatrix} \int_{0}^{\Delta t} \frac{(1-e^{-\beta_{x}\tau})^{2}}{\beta_{x}^{2}} \sigma_{x}^{2} d\tau & \int_{0}^{\Delta t} \frac{(1-e^{-\beta_{x}\tau})e^{-\beta_{x}\tau}}{\beta_{x}} \sigma_{x}^{2} d\tau\\  \int_{0}^{\Delta t} \frac{(1-e^{-\beta_{x}\tau})e^{-\beta_{x}\tau}}{\beta_{x}} \sigma_{x}^{2} d\tau &  \int_{0}^{\Delta t} (e^{-\beta_{x}\tau})^{2}\sigma_{x}^{2} d\tau \end{bmatrix},
\end{equation}
\begin{equation}\label{state4}
Q_{z}=\frac{\sigma_z^{2}}{2\beta_{z}}(1-e^{-2\beta_{z} \Delta t}).
\end{equation} 

\begin{figure}[httb]
	\centering
	\includegraphics[width=0.85\columnwidth]{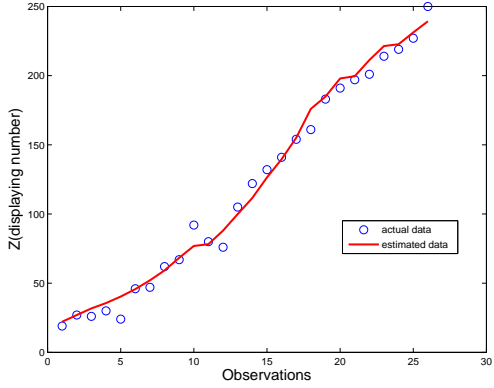}
	\caption{The actual observation data vs. the calculated data with the receiving model.}
	\label{fig:zavszt}
\end{figure}

The second part is the receiving system, the receiving unit displays the received power with the integers between 0 and 255, we use the following equation to link the state vector to the measurement data \cite{janaswamy_bird_1},
 \begin{equation} \label{mea1}
 tanh^{-1}(\frac{Z-Z_{m}}{Z_{M}-Z_{m}})=b\cdot ln(\frac{\xi ^{2}}{P_{0}}+1),
 \end{equation}
$Z_{m}=0$ and $Z_{M}=255$ denote the bounds of the displaying integers, $Z$ is the displaying number which is also the measurement we are given, $\xi ^{2}$ is the received power in the noiseless environment which is calculated by the radiation pattern of the receiving antenna and the relative position of the target and the receiver, $P_{0}$ can be seen as the noise part in the environment.
The values of $b$ and $P_{0}$ are solved by the testing data of a kite via the nonlinear least squares algorithm, the actual time-indexed position of the target is known, the comparison between the actual measurement data and the calculated numbers by the proposed receiving model is plottedn in Fig. \ref{fig:zavszt}, it is clear that the red line (estimated results) matches the blue dots (actual measurements) very well with $b=0.3012$, $P_{0}=4.3458\times 10^{-11}$.

The measurement model is nonlinear since the measurement ($Z$) is a highly nonlinear function with respect of the state vector in equation \ref{mea1}. In the general Kalman filter, we keep the first order term with Taylor series expansion with getting rid of higher order terms to linearize the nonlinear model, and the expectation and variance of the measurement is calculated based on the linear term, this helps us find the arrpoximate results quickly but sacrifice the accuracy.

In this paper, we have two objectives, we begin in section \RNum{2} to deal with the nonlinearity of the measurement model with using unscented transformation for offering more accurate approximation of the nonlinear function, the second objective is to estimate the unknown parameters ($\beta_{x}$, $\beta_{y}$, $\beta_{z}$) in system model with the given measurement, which is presented in section \RNum{3}.  In section \RNum{4}, we present some results with using unscented transformation and optimized parameters, the paper is concluded in section \RNum{5}.

\section{Estimating the Trajectory}
Our goal is to track the target, the procedure of estimating the trajectory with combing the state model and measurement model is presented in this section. The following conditions cause it challenging to estimate the accurate position of the target: many potential positions correspond to one given measurement, the measurement is highly nonlinear with the target's position, the receiving system is working at the noisy environment and several unknown parameters exist in movement system (Sec.~\RNum{3}). As we know, using Kalman filter, the mean value of the estimated positions is generated by the movement system and the receiving system with the given measurement data this technology has been used widely in tracking the target, and it has good accuracy in the noisy environment.
The extended Kalman filter (EKF) \cite{Dan1} is used in \cite{janaswamy_bird_1} following the linearization step for the received power, the limitations of using extended Kalman filter include the inaccuracy and inefficiency, the inaccuracy arises in using Taylor expansion in linearization, only the first order term in approximation equation is kept with getting rid of higher order terms, the inefficiency here refers to the complexity of getting the derivative of the received signal power with respect to the state variables.
Here we use the unscented transformation \cite{Simon1} to obtain signal power's approximate value which will be used in the estimating step to improve the two aspects.

Equation \ref{mea1} is applied to link the measurement to the power of the signal in the noisy environment, we begin in figuring out the relationship between the power and the state vector of the target, the power of the received signal is written as 
 \begin{equation}
  h(\widehat{X})=(\xi(\bar{X}) + \sqrt{P_{0}} \gamma) ^2
 \end{equation}
 where the value of $P_{0}$ has been estimated, $\gamma \sim N(0,1)$ denoting the noise process, the received power consists of two parts, one is from the transmits which depends on the position of the target and the other one is noise which is independent of the target, but the square relates them, $\widehat{X}=[\bar{X}', \gamma]'$ which is a column vector, the noise component is added to the original state variables even though the characteristic of it is known, we do that in order to get the mean and variance with unscented transformation.
The power is nonlinear with $\widehat{X}$, assuming the nominal point is $\widehat{X}_{0}$, relying on Taylor expansion we can get
 \begin{equation} \label{y_true}
   \begin{aligned}
  h(\widehat{X}) \approx  h(\widehat{X}_{0})+\frac{\partial h(\widehat{X})}{\partial \widehat{X}} |_{\widehat{X}_{0}}(\widehat{X}-\widehat{X}_{0})\\ 
  +\frac{1}{2!}(\sum_{t=1}^{n}\frac{\partial h}{\partial \widehat{X}_{t}}|_{\widehat{X}_{0}}(\widehat{X}_{t}-(\widehat{X}_{0})_{t}))^{2} +... \ .
  \end{aligned}
 \end{equation}
 
In Kamlam filter, the expectation and variance of measurement are required to update predicted state variables following the prediction step, to deal with the nonlinearity, only the linear term in the above equation is kept such as the EKF, then the expectation of power is calculated by
  \begin{equation} \label{y_mean}
     \begin{aligned}
      E(h(\widehat{X}))= & h(\bar{X}_{0},\gamma_{0})+E(\frac{\partial h(\widehat{X})}{\partial \bar{X}}|_{(\bar{X}_{0},\gamma_{0})}(\bar{X}-\bar{X_{0}})) \\ 
      &+E(\frac{\partial h(\widehat{X})}{\partial \gamma}|_{(\bar{X}_{0},\gamma_{0})}(\gamma-\gamma_{0})),
     \end{aligned}
 \end{equation}  
in equation \ref{y_mean}, we write the state variables separately to emphasize that only $\bar{X}$ are the variables we want to solve, $\gamma$ is the known noise component.
Assuming $\widehat{X}$ is the predicted state vector at time $t$, the nominal point $\bar{X}_{0}$ is the first term at right hand side of equation \ref{state1} and $\gamma_{0}=0$, $(\bar{X}-\bar{X_{0}})$ is equal to the latter term at right hand side of equation \ref{state1} which is with zero-mean, then the expectation of the power in general EKF is simplified to
 \begin{equation} \label{y_mean_ekf}
E(h(\widehat{X}))= h(\bar{X}_{0},\gamma_{0}).
 \end{equation}  
 
The expectation of the power with unscented transformation is calculated by
  \begin{equation} \label{y_mean_uns}
     E(h(\widehat{X}))=\frac{1}{2n}\sum_{t=1}^{2n}h(\widetilde{X} ^{(t)}),
  \end{equation}  
 \begin{equation} \label{y_mean_uns2}
  \widetilde{X} ^{(t)}=\left\{\begin{matrix}
\widehat{X}_{0}+(\sqrt{nQ})^{T}_{t}, t=1,2...n, \\ 
\widehat{X}_{0}-(\sqrt{nQ})^{T}_{t-n}, t=n+1,n+2...2n,
\end{matrix}\right.
  \end{equation} 
  
\begin{equation}   
Q=\begin{bmatrix}
 \bar{Q}& 0\\ 
 0& 1
\end{bmatrix},\bar{Q}=\begin{bmatrix}
Q_{x} & 0 & 0\\ 
 0 & Q_{y} & 0\\ 
 0& 0 & Q_{z} \\ 
\end{bmatrix},
\end{equation} 
where $n$ = 6 denotes the number of variables, $Q$ is $n$ by $n$ covariance matrix, $Q_{(n,n)}$ denotes the variance of $\gamma$ which is constant, $(\sqrt{\cdot \cdot})_{i}$ is the $i$th row of the square root of certain matrix, expanding equation \ref{y_mean_uns} around the nominal point
\begin{equation} \label{y_mean_uns3}
     \begin{aligned}
     E(h(\widehat{X}))\approx & \frac{1}{2n}\sum_{t=1}^{2n}(h(\bar{X}_{0},\gamma_{0})+\sum_{i=1}^{n}\frac{\partial h}{\partial \widehat{X}_{i}}|_{(\bar{X}_{0},\gamma_{0})}(\widetilde{X}^{(t)}-\widehat{X}_{0})_{i}\\
     &+\frac{1}{2!}(\sum_{i=1}^{n}\frac{\partial h}{\partial \widehat{X}_{i}}|_{(\bar{X}_{0},\gamma_{0})}(\widetilde{X}^{(t)}-\widehat{X}_{0})_{i})^{2}+\cdot \cdot \cdot ),
     \end{aligned}
 \end{equation}  
where $i$ denotes the $i$th element in the vector, from equation \ref{y_mean_uns2}, we can see $(\widetilde{X}^{(t)}-\widehat{X}_{0})_{i}+(\widetilde{X}^{(t+n)}-\widehat{X}_{0})_{i}=0$ for $t=1,2...n$, it implies  $\sum_{t=1}^{2n}(\sum_{i=1}^{n}\frac{\partial h}{\partial \widehat{X}_{i}}|_{(\bar{X}_{0},\gamma_{0})}(\widetilde{X}^{(t)}-\widehat{X}_{0})_{i})^{k}=0$ when $k$ is an odd number, equation \ref{y_mean_uns3} can be simplified to
\begin{equation} \label{y_mean_uns4}
     \begin{aligned}
&E(h(\widehat{X}))\approx  h(\bar{X}_{0},\gamma_{0})+ \frac{1}{4n}\sum_{t=1}^{2n}(\sum_{i=1}^{n}\frac{\partial h}{\partial \widehat{X}_{i}}|_{(\bar{X}_{0},\gamma_{0})}(\widetilde{X}^{(t)}-\\
&\widehat{X}_{0})_{i})^{2}+\frac{1}{2n}\sum_{t=1}^{2n}\frac{1}{4!}(\sum_{i=1}^{n}\frac{\partial h}{\partial \widehat{X}_{i}}|_{(\bar{X}_{0},\gamma_{0})}(\widetilde{X}^{(t)}-\widehat{X}_{0})_{i})^{4}+\cdot \cdot \cdot  ,
     \end{aligned}
 \end{equation}  
combing equation \ref{y_mean_uns2} and \ref{y_mean_uns4}, the expectation can be written as  
\begin{equation} \label{y_mean_uns5}
     \begin{aligned}
&E(h(\widehat{X}))\approx  h(\bar{X}_{0},\gamma_{0})+ \frac{1}{2}\sum_{m=1,k=1}^{n}\frac{\partial^2 h}{\partial \widehat{X}_{m}\partial \widehat{X}_{k}}|_{(\bar{X}_{0},\gamma_{0})}Q_{mk}\\
&+\frac{1}{2n}\sum_{t=1}^{2n}\frac{1}{4!}(\sum_{i=1}^{n}\frac{\partial h}{\partial \widehat{X}_{i}}|_{(\bar{X}_{0},\gamma_{0})}(\widetilde{X}^{(t)}-\widehat{X}_{0})_{i})^{4}+\cdot \cdot \cdot  ,
     \end{aligned}
 \end{equation} 
the above equation is the approximate expectation of received signal power in noisy environment with unscented transformation. As we know, equation \ref{y_true} strands for the most accurate approximate value for the power around the nominal point, we can compare the results with it to see the difference, from equation \ref{y_true} we can get  
\begin{equation} \label{y_true2}
     \begin{aligned}
E(h(\widehat{X}))&=h(\bar{X}_{0},\gamma_{0})+
E(\sum_{i=1}^{n}\frac{\partial h}{\partial \widehat{X}_{i}}|_{\widehat{X}_{0}}(\widehat{X}_{i}-(\widehat{X}_{0})_{i}))\\
&+\frac{1}{2!}E((\sum_{i=1}^{n}\frac{\partial h}{\partial \widehat{X}_{i}}|_{\widehat{X}_{0}}(\widehat{X}_{i}-(\widehat{X}_{0})_{i}))^{2}) \cdot \cdot \cdot \ ,
     \end{aligned}
 \end{equation} 
where $i$ denotes the $i$th element in the vector, the expectation of odd order term can be proved to be equal to 0 because of the symmetric probability density distribution for the component inside parentheses at 0, the above equation can be simplified to     
\begin{equation} \label{y_true3}
     \begin{aligned}
E(h(\widehat{X}))&=h(\bar{X}_{0},\gamma_{0})+
\frac{1}{2}E((\sum_{i=1}^{n}\frac{\partial h}{\partial \widehat{X}_{i}}|_{\widehat{X}_{0}}(\widehat{X}_{i}-(\widehat{X}_{0})_{i}))^{2})\\
&+\frac{1}{4!}E(\sum_{i=1}^{n}\frac{\partial h}{\partial \widehat{X}_{i}}|_{\widehat{X}_{0}}(\widehat{X}_{i}-(\widehat{X}_{0})_{i}))^{4}\cdot \cdot \cdot \ ,
     \end{aligned}
 \end{equation} 
 then we can get 
 \begin{equation} \label{y_true4}
     \begin{aligned}
   E(h(\widehat{X}))&= h(\bar{X_{0}},\gamma_{0})+\frac{1}{2}\sum_{m=1,k=1}^{n}\frac{\partial^2 h}{\partial \widehat{X}_{m}\partial \widehat{X}_{k}}|_{(\bar{X}_{0},\gamma_{0})}Q_{mk}\\
     &+\frac{1}{4!}E(\sum_{t=1}^{n}\frac{\partial h}{\partial \widehat{X}_{t}}|_{\widehat{X}_{0}}(\widehat{X}_{t}-(\widehat{X}_{0})_{t}))^{4} \cdot \cdot \cdot \ ,
     \end{aligned} 
 \end{equation} 
comparing equations \ref{y_mean_ekf} and \ref{y_mean_uns4} with the above equation, we can see more correct components (up to third order) in the expectation expression are kept when using the unscented transformation, thus the accuracy should be improved in term of the expectation. 

The variance of the measurement with unscented transformation is calculated as 
 \begin{equation} \label{variance1}
 \sigma(h(\widehat{X}))=\frac{1}{2n}\sum_{t=1}^{2n}(h(\widetilde{X}^{(t)})-E(h(\widehat{X})))(\cdot \cdot \cdot)^{T},
 \end{equation}
the transpose is not necessary in this situation because the power is a scalar, using Taylor series expansion, we can get
\begin{equation} \label{variance2}
\begin{aligned}
h(\widetilde{X}^{(t)})=&h(\bar{X_{0}},\gamma_{0})+\sum_{i=1}^{n}\frac{\partial h}{\partial \widehat{X}_{i}}|_{\widetilde{X}_{0}}(\widetilde{X}_{i}^{(t)}-(\widehat{X}_{0})_{i})
\\&+\frac{1}{2!}(\sum_{i=1}^{n}\frac{\partial h}{\partial \widehat{X}_{i}}|_{\widehat{X}_{0}}(\widetilde{X}_{i}^{(t)}-(\widehat{X}_{0})_{i}))^{2} \cdot \cdot \cdot \,
\end{aligned}
\end{equation}
 $E(h(\widehat{X}))$ is given in equation \ref{y_mean_uns3} for the unscented transformation, then the variance can be written as
\begin{equation} \label{variance3}
\begin{aligned}
&\sigma{(h(\widehat{X}))}=\frac{1}{2n}\sum_{t=1}^{2n}((\sum_{i=1}^{n}\frac{\partial h}{\partial \widehat{X}_{i}}|_{\widehat{X}_{0}}(\widetilde{X}_{i}^{(t)}-(\widehat{X}_{0})_{i})\\
&+\frac{1}{2!}(\sum_{i=1}^{n}\frac{\partial h}{\partial \widehat{X}_{i}}|_{\widehat{X}_{0}}(\widetilde{X}_{i}^{(t)}-(\widehat{X}_{0})_{i}))^{2} \\
&-\frac{1}{4n}\sum_{m=1}^{2n}(\sum_{i=1}^{n}\frac{\partial h}{\partial \widehat{X}_{i}}|_{(\bar{X}_{0},\gamma_{0})}(\widetilde{X}^{(t)}-\widehat{X}_{0})_{i})^{2}\cdot \cdot \cdot )(\cdot \cdot \cdot )^{T}),
\end{aligned}
\end{equation} 
expand the above equation, for the first product inside parentheses we can get 
\begin{equation} \label{variance4}
\begin{aligned}
&\frac{1}{2n}\sum_{t=1}^{2n}((\sum_{i=1}^{n}\frac{\partial h}{\partial \widehat{X}_{i}}|_{\widehat{X}_{0}}(\widetilde{X}_{i}^{(t)}-(\widehat{X}_{0})_{i})\cdot\\
&(\sum_{i=1}^{n}\frac{\partial h}{\partial \widehat{X}_{i}}|_{\widehat{X}_{0}}(\widetilde{X}_{i}^{(t)}-(\widehat{X}_{0})_{i})^{T})=\frac{\partial h}{\partial \widehat{X}}Q(\frac{\partial h}{\partial \widehat{X}})^{T},
\end{aligned}
\end{equation} 
in the general EKF, the variance of the power only keeps results in equation \ref{variance4}, with using unscented transformation, more terms with correct signs are kept, the true expansion of the variance can be obtained with the expectation of the power given in equation \ref{y_true4}.     

We picked up one point in the map as the potential position and chose one antenna as the receiver to test the validity of unscented transformation, the state vector we set is [410000; $v_{x}$; 4602500; $v_{y}$; 5.4], the velocities do not affect the power, you can assign any value to them, the variances we assigned for $p_{(x,y,z)}$  are 2000, 5000 and 1 with assuming the covariance is 0, in addition $\gamma \sim N(0,1)$.
10000 samples are generated according to the conditions we set, we can calculate the expectation and variance of the power by the measurement equation with the 10000 samples, the result should be close to true values.
The expectation and variance with unscented transformation and general linearization are also presented in Table~\ref{mean1}.   

\begin{table}[H]
\centering

\renewcommand{\arraystretch}{1.2}
\begin{tabular}{|p{3cm}|p{2cm}|p{2cm}|}
\hline
                                     & mean       & variance   \\ \hline
True(10000 samples)   & 6.2054e-11 & 7.6829e-20 \\ \hline
Unscented transformation            & 6.5007e-11 & 1.4493e-20 \\ \hline
General linearization & 1.3402e-11 & 8.3374e-21 \\ \hline
\end{tabular}
\caption{The approximate expectation and variance} \label{mean1}
\end{table}
 
We can see the expectation and variance with unscented transformation are both closer to the true ones at the point we set compared with the results obtained by keeping only the linear terms. 
 
To estimate the trajectory of the target, we begin in guessing the initial state variables and covariance($\bar {X}(0)^{+}$ , $\bar{P}(0)^{+}$), assuming the nominal point is ($\bar{X}(k)^{-};0$), the state variables will be updated by Kalman filter and measurement in the recursive way ($k$=1,2,3,....)
\begin{equation} \label{ukf1}
  \bar {X}(k)^{-}=H(k)\cdot \bar {X}(k-1)^{+},
\end{equation} 
\begin{equation} \label{ukf2}
  \bar{P}(k)^{-}=H(k)\bar{P}(k-1)^{+}H(k)'+\bar{Q}(k),
\end{equation} 
\begin{equation}  \label{ukf3}
  \widehat{\bar{X}}(k)^{-}=[(\bar{X}(k)^{-})^{'}, \bar{\gamma}=0]',
\end{equation} 
 \begin{equation} \label{ukf4}
  \widetilde{X}(k)_{i}=\left\{\begin{matrix}
\sqrt{n\widehat{P}(k)}^{T}_{i}\\ 
-\sqrt{n\widehat{P}(k)}^{T}_{i}
\end{matrix}\right.,
\widehat{P}(k)=\begin{bmatrix}
  \bar{P}(k)^{-} & 0\\ 
 0& 1
\end{bmatrix},
\end{equation} 
\begin{equation} \label{ukf5}
  \bar{Y}(k)=\frac{1}{2n}\sum_{i=1}^{2n}h(  \widehat{\bar{X}}(k)^{-}+  \widetilde{X}(k)_{i}),
\end{equation}  
\begin{equation}  \label{ukf6}
  v(k)=Y(k)-\bar{Y}(k),
\end{equation}  
 \begin{equation} \label{ukf7}
  \begin{aligned}
  F(k)=\frac{1}{2n}\sum_{i=1}^{2n}&((h(  \widehat{\bar{X}}(k)^{-}+  \widetilde{X}(k)_{i})-\bar{Y}(k))\cdot \\
  &(h(  \widehat{\bar{X}}(k)^{-}+  \widetilde{X}(k)_{i})-\bar{Y}(k))'),
   \end{aligned}
\end{equation}  
\begin{equation} \label{ukf8}
  P_{xy}=\frac{1}{2n}\sum_{i=1}^{2n}(\widetilde{X}(k)_{i})(h(  \widehat{\bar{X}}(k)^{-}+  \widetilde{X}(k)_{i})- \bar{Y}(k)),
\end{equation}
\begin{equation} \label{ukf9}
  M(k)=P_{xy}(F(k))^{-1},
\end{equation} 
 \begin{equation} \label{ukf10}
  \bar{X}(k)^{+}=\bar{X}(k)^{-}+M(k)v(k),
\end{equation} 
\begin{equation} \label{ukf11}
  \bar{P}(k)^{+}=\bar{P}(k)^{-}-M(k)F(k)M(k)',
\end{equation} 
in practice, we only keep the first $(n-1)$ elements of the vector $M(k)$ in equation \ref{ukf9}, and apply it to the following equations, because the last variable is noise component which is not necessary to be updated.

\section{Parameters Estimation}
In the movement model, we use $\beta_{(x,y,z)}$ as the parameters to govern the motion of the target, it is necessary to know or choose the appropriate values for these parameters prior to using Kalman filter to estimate the trajectory of the target.
From the system model and Kalman filter, we can see the value of $\beta_{(x,y,z)}$ determines the prediction of the state vector in Kalman filter directly, it is intuitive to set the criterion for the good values of $\beta_{(x,y,z)}$ as making the prediction of the state vector approach the true one, considering the state vector in this project can not be observed directly, we replace the state vector with the measurement data in the criterion, to be more specific, we use the power of the signal.
In equation \ref{state1}, we can see the prediction of the state vector at the next moment is represented by two terms, the former one is deterministic with the given interval and $\beta_{(x,y,z)}$, the latter term consists of variables which are subject to normal distribution, both of the two term are related to $\beta_{(x,y,z)}$, in other words, the distribution of the prediction of the state vector depends on $\beta_{(x,y,z)}$. 
As it is mentioned earlier, the measurement depends on the state vector nonlinearly, thus the measurement should also be the variable which is subject to certain distribution and the characteristic of the distribution is related to $\beta_{(x,y,z)}$. Now the problem becomes finding the desired values for the parameters with given measurements and the distribution of the prediction of the measurement is related to the parameters.
As we know, the maximum likelihood estimation is widely used in estimating the unknown parameters, the basic idea of maximum likelihood estimation is to find the appropriate values for the parameters that maximize the joint probability for the given observations, in other words, when the likelihood function's value is higher, it means the more likely you can get the given observations with the corresponding parameters. 

The general procedure of estimating the unknown parameters with given observations is: 1) calculate the probability for the single observation with the initial guess of the parameters, 2) calculate the joint probability of the entire observations, 3) adjust the values of the parameters to increase the joint probability, 4) if the results reach the requirements, quit.

\begin{figure}[httb]
	\centering
	\includegraphics[width=0.6\columnwidth]{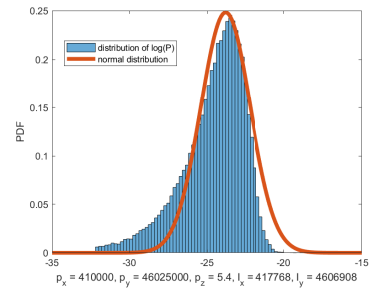}
	\caption{distribution of $\log$ P vs. normal distribution}
	\label{fig:lognormaldist}
\end{figure}

In this paper, the observation is the displaying number $Z(k)$, we convert it to the power of the signal $Y(t)$ according to equation \ref{mea1} first, the expectation and variance of the prediction of the power $\widehat{Y}(k)$ are derived in preceding section in equation \ref{ukf5} and \ref{ukf7}, the high nonlinearity of the power  with respect to state vector causes it hard to obtain the exact and complete expression of the distribution of $\widehat{Y}(k)$ from the distribution of $\widehat{X}(k)^{-}$, we approximate the distribution of $\widehat{Y}(k)$ to log-normal distribution via experiment.
In each experiment, 10000 samples of the state variables are generated with given mean and variance, the distribution of the logarithm of received signal strength and the normal distribution for one example is presented in \ref{fig:lognormaldist}.
The likelihood function of the logarithm of measurement $Y(k)$ is 
\begin{equation} \label{eqt:lognormalPDF}
L(\ln Y (k) | \mu (k), \sigma(k)^{2}) = \frac{1}{\sigma(k) \sqrt{2\pi}}exp(-\frac{\left[\ln Y(k)-\mu(k) \right]^{2}}{2\sigma(k)^{2}}), 
\end{equation}
assuming the observations are independent with each other, the joint likelihood is
\begin{equation} \label{eqt:lognormalJointPDF}
L(\ln Y|\mu,\sigma)=\prod_{k=1}^{n}L(\ln Y (k) | \mu (k), \sigma(k)^{2}),
\end{equation} 
where $\mu(k) = \ln(\frac{\bar{Y}(k)}{\sqrt{T}})$,$ \sigma(k)^{2} = \ln(T)$, $T =1 + \frac{F(k)}{\bar{Y}(k)^{2}}$, $\mu(k)$ and $\sigma(k)^{2}$ are both functions of $\beta_{(x,y,z)}$,
rewrite the above equation to logarithmic format for convenience and denoting $\ln Y(k)-\mu(k)=v(k)$,
\begin{equation} \label{mle2}
  L(\beta_{(x,y,z)})=-\frac{n}{2}\ln (2\pi)-\sum_{k=1}^{n}\ln \sigma (k)-\frac{1}{2}\sum_{k=1}^{n}\frac{v(k)^{2}}{\sigma(k)^{2}},
\end{equation} 
since the first term is constant, our goal becomes finding out $\beta_{(x,y,z)}$ that maximize the value of latter two terms ($\underset{\beta_{(x,y,z)}}{argmax}(-\sum_{k=1}^{n}\ln \sigma(k)-\frac{1}{2}\sum_{k=1}^{n}\frac{v(k)^{2}}{\sigma (k)^{2}})$) or minimize the value of latter two terms without negative sign ($\underset{\beta_{(x,y,z)}}{argmin}(\sum_{k=1}^{n}\ln \sigma (k)+\frac{1}{2}\sum_{k=1}^{n}\frac{v(k)^{2}}{\sigma (k)^{2}})$), we choose the latter one and it is called negative log likelihood, $\beta_{(x,y,z)}$ are require to be positive in order to keep stability of the movement model, we transform them to $\phi_{(x,y,z)}=log(\beta_{(x,y,z)})$ to avoid applying the additional constraints to them in the following optimization step. Until now, we convert the estimation problem into a problem of finding the minimum of multivariable function. 
There are many algorithms available for solving this kind of problem, we tried two algorithms to obtain the desired values.
The Newton's method is equaling the partial derivative of the likelihood function with respect to the unknown variables to 0 to make the variables update towards the desired direction 
\begin{equation} \label{mle3}
  \frac{\partial L}{\partial \beta_{x}}=0, \frac{\partial L}{ \partial \beta_{y}}=0, \frac{\partial L}{\partial \beta_{z}}=0,
\end{equation} 
another algorithm we used is the particle swarm optimization (PSO) algorithm, the cost function is certain negative likelihood function and the particles are the unknown variables, at each iteration the position of the particle is updated according to the history global optimal value of the cost function and the current value of the cost function, comparing with Newton's method, PSO algorithm is simpler to implement but less efficient that takes more iterations. The algorithm of using PSO for estimating parameters is presented and the estimated trajectory with using the results of the two methods are presented in section \RNum{4}.

\begin{algorithm} 
 \caption{Estimating Parameters with PSO}
       \begin{algorithmic}[1] 
        \For{each $i\in [1,m]$}  
          \State initialize the particles $\phi_{i}$ with random number;  
          \State calculate the cost function value $L_{i}$ for each dimension; 
          \State update the best cost function value $L_{gb}$ and the corresponding particles' value $\phi_{gb}$; 
        \EndFor   
        \For{$k=1$; $k<n$; $k++$ }  
         \For{each $i\in [1,m]$}  
               \State update the velocity $v_{i}$ of the particles;   
               \State update the position $\phi_{i}$ of the particles;
                 \If{$L_{i}<L_{gb}$}  
                    \State $L_{gb}=L_{i}$;
                    \State $\phi_{gb}=\phi_{i}$;
                 \EndIf
            \EndFor  
    \EndFor  
    \State \Return{$L_{gb}$ and $\phi_{gb}$}
     
        \end{algorithmic}

\end{algorithm}

\section{Results}
\begin{figure}[httb]
	\centering
	\includegraphics[width=0.75\columnwidth]{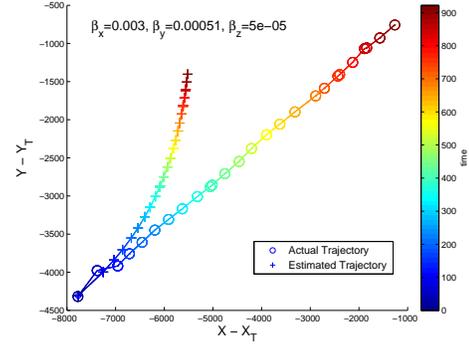}
	\caption{Actual trajectory vs. estimated trajectory with random parameters}
	\label{initial}
\end{figure}

We used one kite to simulate the trajectory of the bird, the GPS device is installed in the kite to obtain the accurate spatial data of the kite, the height of the kite is kept to be around 30 meters, the displaying number of receiving unit and detection time are recorded for estimating the trajectory.
First of all, we used Newton's method and PSO to estimate the values of the parameters in EKF, the initial guess we set for $\beta_{(x,y,z)}$ is [$3*10^{-3},5.1*10^{-4},5*10^{-5}$], the comparison between the true trajectory of the kite and the estimated trajectory with the initial values of $\beta_{(x,y,z)}$ is shown in Fig. \ref{initial}, the true trajectory is denoted by `$\circ$' while the estimated one is plotted with `+'.
The $XY$-axis denotes the relative position of the target with the location of the antenna which detected the target at first.
We can see the estimated trajectory differs from the true one in the flying direction, only the first several points match the actual trajectory which depends primarily on the initial position we set.   

\begin{figure}[httb]
	\centering
	\includegraphics[width=0.75\columnwidth]{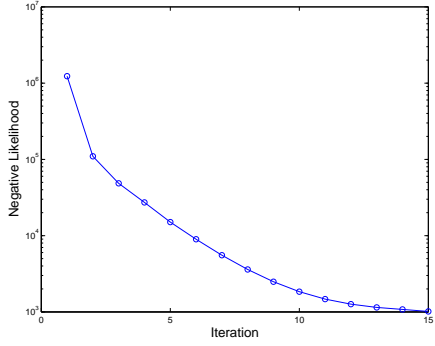}
	\caption{Negative likelihood vs. iteration with Newton's method}
	\label{likelihood_new}
\end{figure}

\begin{figure}[httb]
	\centering
	\includegraphics[width=0.75\columnwidth]{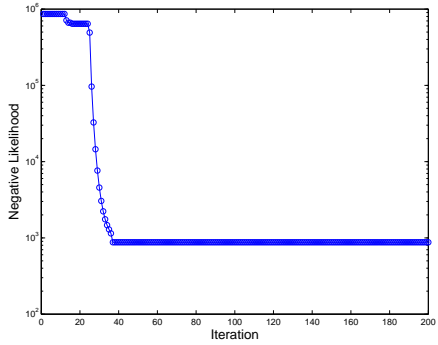}
	\caption{Negative likelihood vs. iteration with PSO method}
	\label{likelihood_pso}
\end{figure}

With using the two algorithms, the negative likelihood vs. iteration are presented in Fig. \ref{likelihood_new} and \ref{likelihood_pso}, the former one is with Newton's method and the latter one is PSO method, the negative likelihood in both methods converged at around 1000 although the PSO needs more iterations.

\begin{figure}[httb]
	\centering
	\includegraphics[width=0.75\columnwidth]{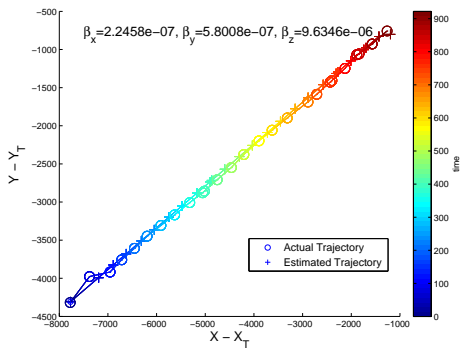}
	\caption{Actual trajectory vs. estimated trajectory with parameters obtained via Newton's method}
	\label{ekfnewton}
\end{figure}

\begin{figure}[httb]
	\centering
	\includegraphics[width=0.75\columnwidth]{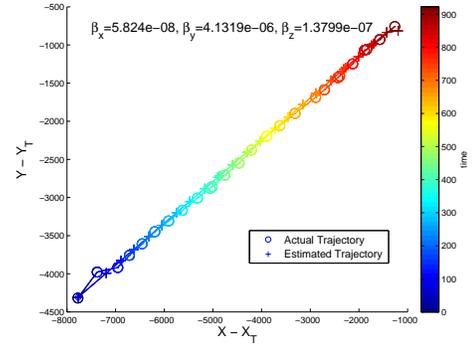}
	\caption{Actual trajectory vs. estimated trajectory with parameters obtained via PSO method}
	\label{ekfpso}
\end{figure}

The parameters $\beta_{(x,y,z)}$ after the estimation procedure are [$2.25*10^{-7},5.8*10^{-7},9.63*10^{-6}$] and [$5.82*10^{-8},4.13*10^{-6},1.38*10^{-7}$] in EKF with Newton's method and PSO method, respectively.
The estimated trajectories with optimized parameters are shown in Fig. \ref{ekfnewton} and  Fig. \ref{ekfpso}. 
Compared with the estimated trajectory in Fig. \ref{initial}, it is clear that not only the direction of the estimated trajectory matches well with the actual one but also the accuracy is improved. 

\begin{figure}[httb]
	\centering
	\includegraphics[width=0.75\columnwidth]{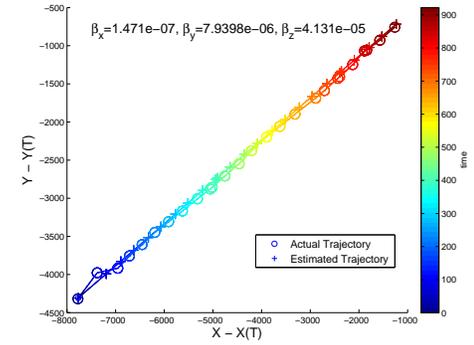}
	\caption{Actual trajectory vs. estimated trajectory with parameters obtained via Newton's method in unscented Kalman filter}
	\label{ukfnewton}
\end{figure}

\begin{figure}[httb]
	\centering
	\includegraphics[width=0.75\columnwidth]{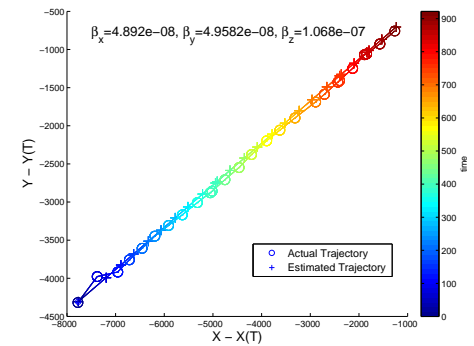}
	\caption{Actual trajectory vs. estimated trajectory with parameters obtained via PSO method in unscented Kalman filter}
	\label{ukfpso}
\end{figure}

We also applied the maximum estimation to Kalman filter with unscented transformation, the parameters $\beta_{(x,y,z)}$ are  [$1.47*10^{-7},7.94*10^{-6},4.13*10^{-5}$] and [$4.89*10^{-8},4.96*10^{-8},1.07*10^{-7}$] through the estimation. 
Fig. \ref{ukfnewton} and Fig. \ref{ukfpso} show the estimated trajectories with using unscented transformation in Kalman filter with estimated $\beta_{(x,y,z)}$, with using unscented transformation, the estimated trajectories also match the actual one well, and here we define 
\begin{equation}
\varepsilon = \sum_{i=1}^{n}[(x_{kite}(i)-x_{p}(i))^{2}+(y_{kite}(i)-y_{p}(i))^{2}],
\end{equation}
where $(x,y)_{kite}$ denotes the true location of the kite, $(x,y)_{p}$ is the estimated location, $\varepsilon=4.16\times10^{5}$ in Fig. \ref{ukfnewton}, while in Fig. \ref{ekfnewton}, $\varepsilon=7.8\times10^{5}$, it means the estimated trajectory with unscented transformation is closer to the true one in this situation. The improvement of using unscented Kalman filter is also demonstrated by obtaining smaller square error with using the same parameters in EKF.  


\section{Conclusion}
In this paper, we use the state space method to estimate the trajectory of the target with the time-indexed signal strength from the telemetry towers. In the movement system, the values of unknown parameters are estimated via the maximum likelihood function method, and the unscented transformation is applied in using Kalman filter to obtain better performance for the accuracy and avoid the linearization step in the general Kalman filter. 
\section{Future work}
The transition equations in the system model is not complete yet, because the external process is not taken into account, we will continue to figure it out, and the initialization plays an important role in estimating the trajectory, we will study how to obtain to optimal guess by the given measurement data.

\ifCLASSOPTIONcaptionsoff
  \newpage
\fi



%


\bibliography{E:/latex_biblio/huabiblio}

\begin{thebibliography}{1}
\providecommand{\url}[1]{#1}
\csname url@samestyle\endcsname
\providecommand{\newblock}{\relax}
\providecommand{\bibinfo}[2]{#2}
\providecommand{\BIBentrySTDinterwordspacing}{\spaceskip=0pt\relax}
\providecommand{\BIBentryALTinterwordstretchfactor}{4}
\providecommand{\BIBentryALTinterwordspacing}{\spaceskip=\fontdimen2\font plus
\BIBentryALTinterwordstretchfactor\fontdimen3\font minus
  \fontdimen4\font\relax}
\providecommand{\BIBforeignlanguage}[2]{{%
\expandafter\ifx\csname l@#1\endcsname\relax
\typeout{** WARNING: IEEEtran.bst: No hyphenation pattern has been}%
\typeout{** loaded for the language `#1'. Using the pattern for}%
\typeout{** the default language instead.}%
\else
\language=\csname l@#1\endcsname
\fi
#2}}
\providecommand{\BIBdecl}{\relax}
\BIBdecl

\bibitem{Durbin1}
J.~Durbin and S.~J. Koopman, \emph{Time Series Analysis by State Space
  Method}.\hskip 1em plus 0.5em minus 0.4em\relax New York: Oxford University
  Press, 2001.

\bibitem{Patterson1}
T.~A. Patterson, ``State space models of individual animal movement,''
  \emph{Trends Ecol Evol.}, pp. 87--94, February 2008.

\bibitem{janaswamy_bird_1}
R.~Janaswamy, P.~Loring, and J.~D. McLaren, ``A state space technique for
  wildlife position estimation using non-simultaneous signal strength
  measurements,'' \emph{arXiv 1805.11171}, 2018.

\bibitem{Dan1}
D.~Simon, \emph{Optimal State Estimation}.\hskip 1em plus 0.5em minus
  0.4em\relax New York: John Wiley Sons, 2006.

\bibitem{Simon1}
S.~Julier, J.~Uhlmann, and H.~F. Durrant-Whyte, ``A new method for the
  nonlinear transformation of means and covariances in filters and
  estimators,'' \emph{IEEE Trans. on Auto. Control}, vol.~45, no.~3, pp.
  477--482, March 2000.

\end{thebibliography}
\bibliographystyle{IEEEtran}

\end{document}